\documentclass[%
 reprint,
superscriptaddress,
nofootinbib,
 amsmath,amssymb,
 aps,
prl,
]{revtex4-2}

\usepackage{graphicx}
\usepackage{dcolumn}
\usepackage{bm}
\usepackage{upgreek}
\usepackage{dcolumn}
\usepackage[normalem]{ulem}
\usepackage{tabularx}
\usepackage{booktabs}
\usepackage{nicefrac}
\usepackage{xcolor}
\usepackage{color}

\definecolor{mbscolor}{rgb}{0.60, 0.0, 0.65}

\newcommand{\mycomment}[1]{}
\newcommand{\bvec}[1]{{\mathbf{#1}}}
\newcommand{\ra}[1]{\renewcommand{\arraystretch}{#1}}
\newcommand{\tls}{\addlinespace[.6em]}

\begin{document}

\preprint{APS/123-QED}

\title{Fingerprints of triaxiality in the charge radii of neutron-rich Ruthenium}

\author{Bernhard Maass}%
\email{maass@anl.gov}
\affiliation{%
Physics Division, Argonne National Laboratory, 
Argonne, Illinois 60439, USA
}%
\affiliation{%
Institut f\"ur Kernphysik, TU Darmstadt, 
64289 Darmstadt, Germany
}%

\author{Wouter Ryssens}
\email{wouter.ryssens@ulb.be}
\affiliation{%
Institut d’Astronomie et d’Astrophysique, Université libre de Bruxelles, 
Campus de la Plaine CP 226, 1050 Brussels, Belgium
}
\affiliation{%
Brussels Laboratory of the Universe - BLU-ULB, Campus de la Plaine CP 226, 1050 Brussels, Belgium}

\author{Kristian K\"onig}%
\affiliation{%
Institut f\"ur Kernphysik, TU Darmstadt, 
64289 Darmstadt, Germany
}%

\author{Michael Bender}
\affiliation{%
Université Claude Bernard Lyon 1, CNRS/IN2P3, IP2I Lyon, 
UMR 5822, Villeurbanne 69622, France
}

\author{Daniel P. Burdette}
\affiliation{%
Physics Division, Argonne National Laboratory, 
Argonne, Illinois 60439, USA
}%

\author{Jason Clark}
\affiliation{%
Physics Division, Argonne National Laboratory, 
Argonne, Illinois 60439, USA
}%

\author{Adam Dockery}
\affiliation{%
Facility for Rare Isotope Beams, Michigan State University, 
East Lansing, Michigan 48824, USA
}%
\affiliation{%
Department of Physics and Astronomy, 
Michigan State University, 
East Lansing, Michigan 48824, USA
}%

\author{Guilherme Grams}
\affiliation{%
Institut d’Astronomie et d’Astrophysique, Université libre de Bruxelles, 
Campus de la Plaine CP 226, 1050 Brussels, Belgium
}

\author{Max Horst}
\affiliation{%
Institut f\"ur Kernphysik, TU Darmstadt, 
64289 Darmstadt, Germany
}%

\author{Phillip Imgram}
\affiliation{%
Institut f\"ur Kernphysik, TU Darmstadt, 
64289 Darmstadt, Germany
}%

\author{Kei Minamisono}
\affiliation{%
Facility for Rare Isotope Beams, Michigan State University, 
East Lansing, Michigan 48824, USA
}%
\affiliation{%
Department of Physics and Astronomy, 
Michigan State University, 
East Lansing, Michigan 48824, USA
}%

\author{Patrick Müller}
\affiliation{%
Institut f\"ur Kernphysik, TU Darmstadt, 
64289 Darmstadt, Germany
}%

\author{Peter M\"uller}%
\affiliation{%
Physics Division, Argonne National Laboratory, 
Argonne, Illinois 60439, USA
}%

\author{Wilfried N\"ortersh\"auser}%
\affiliation{%
Institut f\"ur Kernphysik, TU Darmstadt, 64289 Darmstadt, Germany
}%

\author{Skyy V. Pineda}
\affiliation{%
Facility for Rare Isotope Beams, Michigan State University, 
East Lansing, Michigan 48824, USA
}%
\affiliation{%
Department of Chemistry, 
Michigan State University, 
East Lansing, Michigan 48824, USA
}

\author{Simon Rausch}
\affiliation{%
Institut f\"ur Kernphysik, TU Darmstadt, 
64289 Darmstadt, Germany
}%

\author{Laura Renth}
\affiliation{%
Institut f\"ur Kernphysik, TU Darmstadt, 
64289 Darmstadt, Germany
}%

\author{Brooke J. Rickey}
\affiliation{%
Facility for Rare Isotope Beams, Michigan State University, 
East Lansing, Michigan 48824, USA
}%
\affiliation{%
Department of Physics and Astronomy, 
Michigan State University, 
East Lansing, Michigan 48824, USA
}%

\author{Daniel Santiago-Gonzalez}
\affiliation{%
Physics Division, Argonne National Laboratory, 
Argonne, Illinois 60439, USA
}%

\author{Guy Savard}
\affiliation{%
Physics Division, Argonne National Laboratory, 
Argonne, Illinois 60439, USA
}%

\author{Felix Sommer}
\affiliation{%
Institut f\"ur Kernphysik, TU Darmstadt, 
64289 Darmstadt, Germany
}%

\author{Adrian A. Valverde}
\affiliation{%
Physics Division, Argonne National Laboratory, 
Argonne, Illinois 60439, USA
}%
\noaffiliation

\begin{abstract}
We present the first measurements with a new collinear laser spectroscopy setup at the Argonne Tandem Linac Accelerator System utilizing its unique capability to deliver neutron-rich refractory metal isotopes produced by the spontaneous fission of $^{252}$Cf. We measured isotope shifts from optical spectra for nine radioactive ruthenium isotopes $^{106-114}$Ru, reaching deep into the mid-shell region. The extracted charge radii are in excellent agreement with predictions from the Brussels-Skyrme-on-a-Grid models that account for the triaxial deformation of nuclear ground states. We show that triaxial deformation impacts charge radii in models that feature shell effects, in contrast to what could be concluded from a liquid drop analysis. This indicates that this exotic type of deformation should not be neglected in regions where it is known to occur, even if its presence cannot be unambiguously inferred through laser spectroscopy.

\end{abstract}

\maketitle
\paragraph{\label{sec:introduction}Introduction}

Deformation is a fundamental concept when describing nuclear ground and excited states: an atomic nucleus does not need to be spherically symmetric but can take a range of different shapes in its intrinsic frame. The fingerprint of deformation and its variation with neutron number $N$ and proton number $Z$ can be discerned in numerous nuclear observables, such as masses and radii.
A prominent example is the root-mean-square (rms) nuclear charge radius $R_c = \langle r_c^2 \rangle^{1/2}$. Laser spectroscopy techniques of short-lived radioactive isotopes have uncovered numerous cases where this observable varies rapidly with $N$ along an isotopic chain~\cite{Bonn.1972,thibault1981,buchinger1990,cheal2007a,cubiss2018,Sels2019,barzakh2021}. 
Even if $R_c$ is, in principle, affected by the deformation of all multipolarities, this observable does not carry any information on the angular dependence of the nuclear density and cannot - at least, by itself - be used to pin down which multipole moments are relevant to describe a given nucleus. The vast majority of experimental data on $R_c$ is thus interpreted by limiting the nucleus to axially symmetric shapes whose deviation from sphericity can be measured through a single (dimensionless) multipole moment $\beta_{20}$.
This assumption is usually justified: axially symmetric configurations suffice to describe most nuclei, and the higher order moments of nuclear ground states tend to be significantly smaller than their quadrupole deformation~\cite{Scamps.2021}.

Yet there are several regions of the nuclear chart where triaxial deformation occurs, \textit{i.e.}, where the nuclear density does not have any rotational symmetry axis and has to be characterized by \emph{two} quadrupole multipole moments: either $(\beta_{20}$, $\beta_{22})$ or the more widely used $(\beta_{2}, \gamma)$,
see the supplemental material for their definition. The experimental fingerprint of triaxial deformation is faint: While the analysis of rotational bands can offer indications~\cite{rowe2010a}, it can only be probed directly through Coulomb excitations~\cite{cline1986}.
To our knowledge, there has been no experimental search for the impact of triaxial deformation on the rms charge radius. What literature exists on the subject is limited to the statement that $R_c$ is largely insensitive to $\gamma$; a stance justified by arguments from a simple liquid drop picture~\cite{grechukhin1960,grechukhin,hilberath1992}.

In this Letter, we challenge this state of affairs by presenting the first values for the charge radii of nine neutron-rich ruthenium isotopes, $^{106-114}$Ru. The triaxial deformation of these isotopes - as well as that of the stable $^{102,104}$Ru - is both established experimentally~\cite{garrett2022,esmaylzadeh2022,summerer1980,aysto1990, shannon1994,doherty2017}
and supported by numerous models of different 
origins~\cite{moller2006,moller2008,Scamps.2021,Amedee,nomura2016,zhang2015}.
However, these refractory metals have never been addressed with high-precision laser spectroscopy: their refractory nature makes it challenging to extract them from ISOL-type targets, requiring a buffer-gas-based extraction technique instead. We present data collected from the first experiment at the new Argonne Tandem Hall Laser Beamline for Atom and Ion Spectroscopy (ATLANTIS) at Argonne National Laboratory (ANL). It was specifically designed to target the neutron-rich isotopes of refractory metals produced by the Californium Rare Isotope Breeder Upgrade (CARIBU)  $^{252}$Cf fission source. We compare the experimental findings to the predictions of the Brussels-Skyrme-on-a-Grid (BSkG) model 
series~\cite{Scamps.2021,Ryssens.2022,Grams.2023,Grams.2025}; these state-of-the-art energy-density functional (EDF) models systematically - i.e. across the entire nuclear chart - allow for triaxial deformation when it is energetically favorable. The four available models reproduce the charge radii data remarkably well, as already demonstrated for other observables~\cite{hukkanen2023,hukkanen2023a,stryjczyk2024}. In these models, which include quantum mechanical shell effects absent in the liquid-drop picture, the triaxial deformation impacts the charge radius $R_c$.
Although the effect is somewhat model-dependent, it cannot be ignored when studying the charge radii of isotopes that are known to exhibit triaxial deformation.


\paragraph{Experimental Method}
We installed ATLANTIS in the low-energy experimental area of the Argonne Tandem Linac Accelerator System (ATLAS) that provides access to CARIBU beams. In CARIBU, neutron-rich fission products from a $^{252}$Cf source are thermalized inside a helium-filled large-volume gas catcher~\cite{Savard.2016}, extracted as singly-charged ions and transported to the end station via radio-frequency (RF) and electrostatic guides \cite{Savard.2008, Savard.2011}.
We performed the first isotope shift measurements on neutron-rich ruthenium and demonstrated very high sensitivity for fluorescence-detection laser spectroscopy, recording the spectrum of the least abundant isotope $^{114}$Ru despite a production rate of only a few tens of ions per second.

\begin{figure}
    \centering
    \includegraphics[width=1\linewidth]{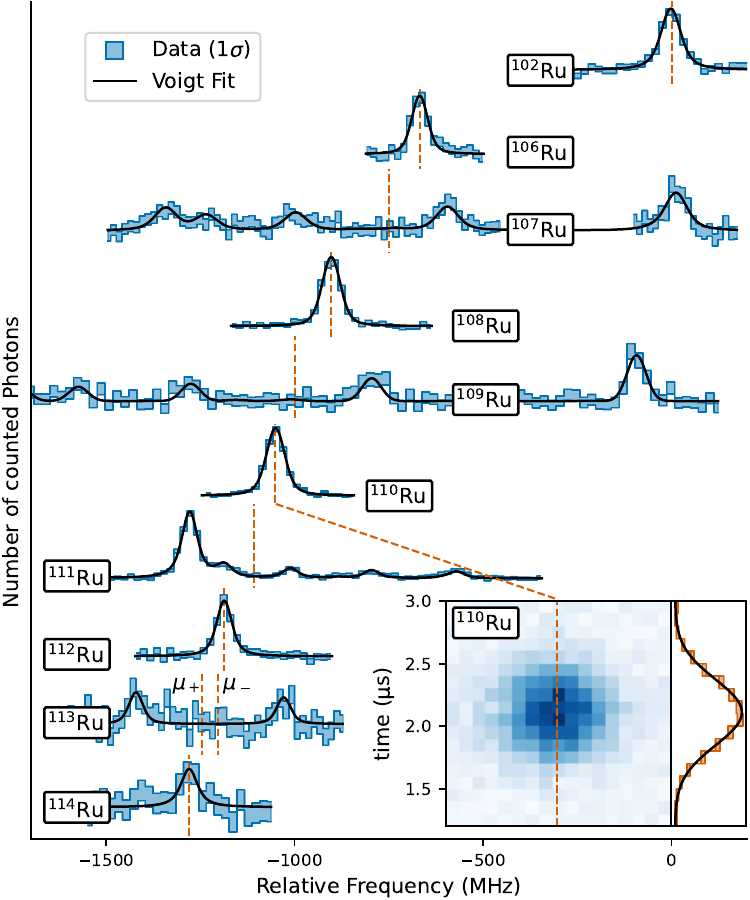}
    \caption{A sample of the recorded data for the stable reference isotope $^{102}$Ru (top row) and all other recorded radioactive isotopes. The solid black lines show a fit to the data. The inset shows a time--frequency profile of a $^{110}$Ru resonance.}
    \label{fig:waterfall}
\end{figure}

The ATLAS low-energy area features a new helium-filled RF quadrupole beam cooler-buncher~\cite{Valverde.2020, Burdette.2025} based on the design described in ~\cite{barquest.2017, Lapierre.2018} . Accumulation times of up to 30\,s were used before releasing bunches with a time spread of $\approx 1\upmu$s. The high compression ratio significantly improves the signal-to-noise ratio of the laser spectroscopic measurement \cite{nieminen.2002}. The temporal signal distribution shows a homogeneous profile which is shown in the inset of Fig.~\ref{fig:waterfall} for $^{110}$Ru. The high voltage applied to the buncher defines the ion starting potential and is actively stabilized to 23\,750\,V via a temperature-stabilized high-precision resistor chain voltage divider \cite{passon2024}. 

A charge-exchange cell, placed directly before the fluorescence detection region, neutralizes the ion beam and allows us to access neutral atoms. In contrast to traditional approaches that use alkaline elements, we use magnesium as the neutralization agent to more closely match the ionization energy of ruthenium. It provided an extraordinarily high charge-exchange efficiency of $\sim90\%$ and a strong g.s. population at approximately $400\,^\circ$C. In addition, we did not observe any noticeable distortion of the lineshapes of the atomic spectra due to the near-resonant charge exchange into the atomic ground state.

With these highly efficient devices in place, we performed laser spectroscopy in the 350-nm ground-state transition of neutral ruthenium between the atomic levels [Kr]\,4$d^7$\,$5s$\:$^5\!F_5$\;(0\,cm$^{-1}$) and [Kr]\,$4d^7$\,$5p$\:$^5\!G_6$\;(28\,571.89\,cm$^{-1}$). The light was generated by a continuous-wave, titanium-sapphire laser (Sirah Matisse) with subsequent frequency doubling in a nonlinear crystal placed in an external enhancement cavity (Spectra Physics Wavetrain). The laser frequency was stabilized to a wavelength meter (High Finesse WS-8) that was referenced to a He:Ne laser with less than 2-MHz frequency drifts within the time frame of typical isotope-shift measurements \cite{Konig2020,Verlinde2020}.

Stable ruthenium ions are obtained from a laser-ablation ion source installed upstream of the buncher.  Each measurement of a radioactive isotope from CARIBU was bracketed by reference measurements on stable $^{102}$Ru to account for potential drifts in voltage or frequency. Care was taken to ensure that online and offline bunch properties were comparable to prevent systematic frequency shifts, particularly those caused by overfilling the buncher for the stable beam.

\paragraph{Data Analysis}
The resonances in the recorded spectra show the best agreement with a Voigt profile, a convolution of Lorentzian ($\gamma$ held fixed at $33(1)$\,MHz) and Gaussian lineshape due to thermal broadening ($\sigma \approx 15$\,MHz, fitted). Figure~\ref{fig:waterfall} shows samples of the recorded spectra. Hyperfine structure (hfs) emerges from the interaction of the electron angular momentum $J$ and the nuclear spin $I$. 
The nuclear magnetic dipole moment and, in case $I > \nicefrac{1}{2}$, the electric quadrupole moment, relative to the known values of $^{99,101}$Ru \cite{Stone2019, Stone2021}, can be extracted from the hfs via the hyperfine parameters $A$ and $B$. The ratios of the hyperfine parameters for the upper and the lower states were fixed to $A_u/A_l=0.4229(3)$ and $B_u/B_l=2.23(2)$, obtained from high precision spectra of the stable isotopes, which are not discussed in this letter but agree with literature \cite{Forest2014}. We can exclude a hyperfine anomaly larger than 1\,\% (2\,\% for $^{113}$Ru).

We fitted the data using a $\chi^2$ optimization package described in \cite{muller2024}.  The uncertainties in centroid frequencies were scaled with the reduced $\chi^2_\mathrm{\nu}$, which was $<2$ for all fits. The obtained values for the isotope shift $\delta\nu^{102,A}$ are listed in Tab.\,\ref{tab:results} with total uncertainties between 3~and~9\,MHz,. These are typically dominated by statistical uncertainty but include systematic errors from the laser wavelength, high-voltage and scan-voltage readout, and $A$/$B$-ratios.

The spectra of $^{113}$Ru show only two peaks, which we attribute to the hfs of an $I=1/2$ nuclear spin as the most likely solution. However, due to large statistical uncertainty and potential interference from isomeric states, we cannot categorically rule out other spin assignments. Additionally, we cannot distinguish between a hfs with a positive or negative magnetic moment, leading to two equally likely solutions for the isotope shift, which differ by $\approx40\,$MHz. Consequently, both cases for $I=1/2$ are presented separately in all tables and plots.

The isotope shift
\begin{equation}
   \delta \nu^{A,A'} = \nu^{A'}-\nu^{A} = \frac{M_{A'}-M_{A}}{(M_{A'}+m_e)(M_{A}+m_e)} K_\infty + F \Lambda^{A,A'}
\end{equation} 
originates from the difference in their nuclear masses $M_A$ plus electron mass $m_e$, scaled with the mass shift parameter $K_\infty$, and their difference in nuclear size, given by the nuclear radius parameter $\Lambda^{A,A'}$, scaled with the field shift parameter $F$~\cite{King.1984}. The nuclear radius parameter to first order can be expressed by the differential root-mean-square charge radius $\delta\left<r_c^2\right>^{A,A'}$ by subtracting 2.5\% higher-order contributions~\cite{Schopper2004} that can not be resolved from isotope shifts in a single spectral line \cite{Papoulia2016}. For this analysis, $\Lambda^{A,A'}$ is obtained from muonic atom spectroscopy of stable isotopes~\cite{Schopper2004}, and the corresponding isotope shifts from a complementary measurement at ATLANTIS, showing excellent agreement with previous experiments~\cite{Forest2014}. With this dataset, we performed a King-plot procedure~\cite{King.1984}, yielding $F=-2.13\,(46)$\,GHz\,fm$^{-2}$ and $K_\infty=954\,(564)$\,u\,GHz. Details can be found in the supplemental material. The results allow us to interpolate and convert the isotope shifts of the newly measured radioactive isotopes to differential charge radii, which can be found in Tab.\,\ref{tab:results}. The uncertainty analysis treats the strongly correlated uncertainties of $F$ and $K_\infty$ appropriately. The King plot interpolation introduces a correlated error to $\delta\left<r_c^2\right>^{102,A}$, denoted in curly brackets in the table, separated from the much smaller measurement uncertainty. The final step is to calculate the absolute charge radii $R_c$ based on the stable reference isotope $^{102}$Ru, $\langle r_c^2 \rangle^{1/2}_{102}=4.481(2)$\,fm~\cite{Schopper2004}. 

\begin{table}
\centering
\ra{1.2}
\caption{The measured isotope shift, differential, and absolute radii of neutron-rich ruthenium. In angle brackets, all uncorrelated statistical and systematic measurement uncertainties are given. The curly brackets denote the correlated uncertainty from the King plot analysis procedure. For the absolute radii, the values in parentheses give their total uncertainty, including that of the reference isotope $^{102}$Ru \cite{Schopper2004}.} 
\begin{tabularx}{0.49\textwidth}{lXD..{4.1}lXD..{1.3}lXD..{1.4}l}
\tls
\toprule[\lightrulewidth]\toprule[\lightrulewidth]

A   && \multicolumn{2}{c}{$\delta\nu^{102,A}$} && \multicolumn{2}{c}{$\delta\left<r_c^2\right>^{102,A}$} && \multicolumn{2}{c}{R$_c$} \\
   && \multicolumn{2}{c}{(MHz)} && \multicolumn{2}{c}{(fm$^2$)} && \multicolumn{2}{c}{(fm)} \\
\midrule
102     &&         &                       &&         &                           && 4.4810   & (20)  \\
106     && -666.7  & $\langle3.5\rangle$   &&  0.490  & $\langle2\rangle$\,\{13\}    && 4.5353   & (24)  \\    
107     && -741.9  & $\langle7.8\rangle$   &&  0.567  & $\langle4\rangle$\,\{15\}    && 4.5438   & (26)  \\
108     && -905.6  & $\langle3.3\rangle$   &&  0.686  & $\langle2\rangle$\,\{17\}    && 4.5569   & (27)  \\    
109     && -1001.8 & $\langle8.8\rangle$   &&  0.771  & $\langle4\rangle$\,\{20\}    && 4.5662   & (29)  \\
110     && -1053.6 & $\langle3.3\rangle$   &&  0.834  & $\langle2\rangle$\,\{26\}    && 4.5731   & (34)  \\
111     && -1112.5 & $\langle5.6\rangle$   &&  0.900  & $\langle3\rangle$\,\{33\}    && 4.5803   & (40)  \\
112     && -1190.1 & $\langle3.5\rangle$   &&  0.974  & $\langle2\rangle$\,\{38\}    && 4.5884   & (45)  \\
113\footnote{evaluated with positive magnetic moment}    && -1245.3 & $\langle4.4\rangle$   &&  1.037  & $\langle2\rangle$\,\{46\}    && 4.5953   & (52)  \\
113\footnote{evaluated with negative magnetic moment}    && -1204.4  & $\langle4.4\rangle$   &&  1.018  & $\langle2\rangle$\,\{49\}    && 4.5932   & (55)  \\
114     && -1282.2  & $\langle6.0\rangle$   &&  1.091  & $\langle3\rangle$\,\{55\}    && 4.6011   & (61)  \\  
\bottomrule[\lightrulewidth]\bottomrule[\lightrulewidth]
\end{tabularx}
\label{tab:results}
\end{table}

\paragraph{Discussion}

\begin{figure*}
    \centering
    \includegraphics[width=1\linewidth]{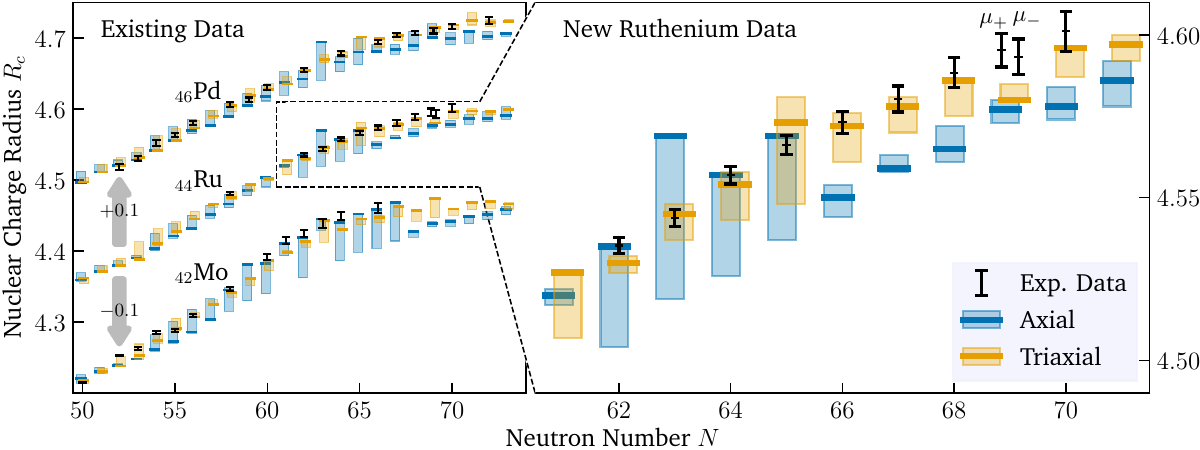}
    \caption{Comparison between the experimental charge radii for the Ru, Mo~\cite{Charlwood.2009}, and Pd~\cite{geldhof2022,Renth.2025} isotopic chains, and BSkG$x$ calculations with the inclusion of triaxial deformation (yellow) and without (blue, see text for details). The colored bands show the spread in predictions between all BSkG$x$, and the solid lines indicate the calculated BSkG4 values.}
    \label{fig:bskg}
\end{figure*}
We compare the new data to the predictions of all models in the Brussels-Skyrme-on-a-Grid series (BSkG1 - BSkG4) that aim to provide nuclear data for astrophysical applications~\cite{kullmann2023,martinet2025a,deprince2025}. The BSkG$x$ are models based on an energy density functional (EDF) of the Skyrme type; they can be treated as similar for the purposes of this study because (i) they all are based on a three-dimensional numerical representation of the nucleus that naturally includes the possibility of triaxial deformation~\cite{Ryssens2016}, (ii) their parameter adjustment protocols - involving data on thousands of nuclei - were nearly identical and (iii) they - globally speaking - describe all known charge radii and the (scarce) Coulomb excitation data on triaxial deformation about equally well~\cite{Scamps.2021,Ryssens.2022,Grams.2023,Grams.2025}. For example, all four models predict $\beta_2 \in [0.264,0.278]$ and $\gamma \in [19^{\circ},23^{\circ}]$ for $^{104}$Ru while experiment indicates $\beta_2=0.272(4), \gamma = 23(3)^{\circ}$~\cite{Srebrny.2006}. They also perform equally well for other nearby stable nuclei with available Coulomb excitation data. Nevertheless, there are important modeling differences between the different BSkG$x$; this is why we take BSkG4 - the most sophisticated entry - as a reference point and use the spread in the calculated charge radii as a proxy for the modeling uncertainty.

In Fig.~\ref{fig:bskg}, we compare two sets of calculations to the experimental data: one set that allows for triaxial deformation (in red) and another set (in blue) that is restricted to axially symmetric configurations with the lowest energy, i.e., either the prolate or the oblate saddle point for each nucleus. The uncertainties are derived from the spread of the four models. We emphasize that we compare \emph{absolute} charge radii; the models require no systematic offset to match the experiment. Focusing first on the Ru isotopes, we see that the triaxial band closely follows the experimental data points across the whole isotopic chain within uncertainties. Only for one of the most exotic isotopes ($N = 69$) do all models significantly underestimate the experimental radii. The axially symmetric configuration performs similarly well for the lighter Ru isotopes but systematically underestimates the charge radii for $N \geq 66$. The axial results also feature a much broader model spread between $ 62 \leq N \leq 65$: there, the calculated charge radii are affected by the transition from prolate to oblate saddle point that occurs at different neutron numbers for each model. The unrestricted, i.e.\ triaxial, BSkG$x$ predictions are clearly better than those from axially symmetric calculations. To show that both the level of agreement and the difference between triaxial and axial calculations is not limited to one isotopic chain, we also compare calculated and measured charge radii for the Mo and Pd chains~\cite{geldhof2022}.
Most of the literature does not consider any possible impact of triaxial deformation on charge radii, even for nuclei that are known to have such asymmetric ground states~\cite{geldhof2022}. We presume that the (mostly unstated) motivation to neglect this degree of freedom is that of Ref.~\cite{hilberath1992}, which states `[...] the isotope shift is very insensitive to triaxial shape.'. This argument is based on the observation that the rms radius of a triaxial quadrupole-deformed liquid drop\footnote{We provide the expression for the radius of such a drop and a more in-depth analysis in the supplementary material.} is almost entirely determined by the size of the quadrupole deformation $\beta_2$ and only marginally sensitive to the triaxiality angle $\gamma$. From this, however, it does not follow that triaxial deformation does not play a role 
for charge radii. A nucleus's energetically most favorable shape is primarily determined by quantum mechanical shell effects~\cite{strutinsky1968}. Since these vary strongly with deformation, changing $\gamma$ will force a rearrangement of the occupation numbers of the filled single-particle levels and their spatial wave functions and thus induce further changes to the spatial density distribution. %
First, the energetically optimal $\beta_2$ will usually not remain fixed when varying $\gamma$. The total quadrupole deformation of a triaxial ground state will typically not match that of the axially symmetric prolate or oblate configuration; even if $\gamma$ would not influence $R_c$ much by itself, the change in $\beta_2$ will. Second, any change of $\beta_2$ or $\gamma$ will induce a change in higher-order moments of the nucleus. Even if often omitted from analyses, these do contribute to radii in liquid drop and EDF-based models alike. For instance, the effect of octupole deformation on the charge radii of Ac isotopes was analyzed in Ref.~\cite{verstraelen2019}. Third, in an EDF calculation, the features of the diffuse radial profile of the charge density also change with $\beta_2$ and $\gamma$; already a small radial redistribution of nucleons far from the nuclear center can then have a sizable effect on the calculated charge radius. While the global effect of nuclear diffuseness is sometimes included in the liquid drop analysis of charge radii \cite{hasse1988,myers1983}, its change with deformation cannot be treated in such models. 

\begin{figure}
    \centering
    \includegraphics[width=0.8\linewidth]{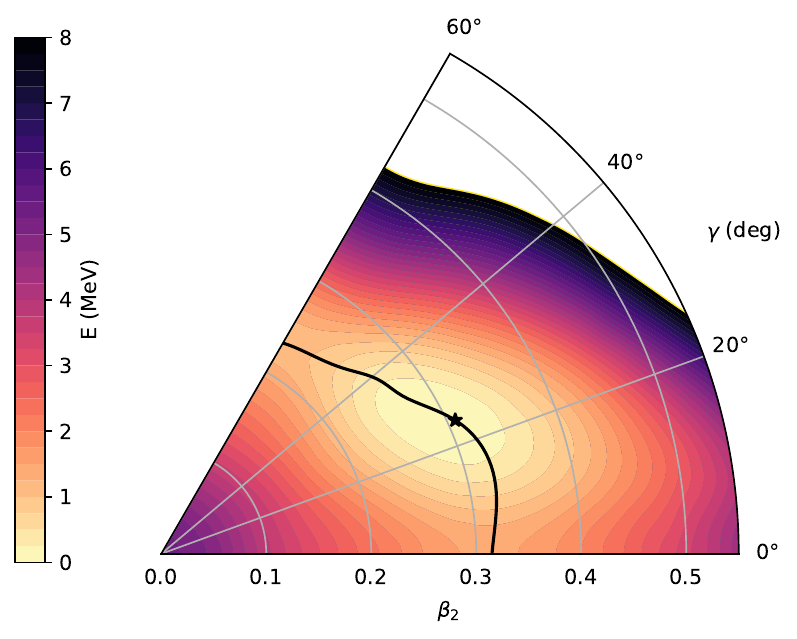}
    \caption{ Normalized total energy of $^{112}$Ru obtained with BSkG4 as a function of $\beta_2$ and gamma.
              The black line indicates the minimal energy as a function of $\gamma$; the global energy minimum is indicated by a black star.}
              \label{fig:pes}
\end{figure}

\begin{figure}
    \centering
    \includegraphics[width=0.8\linewidth]{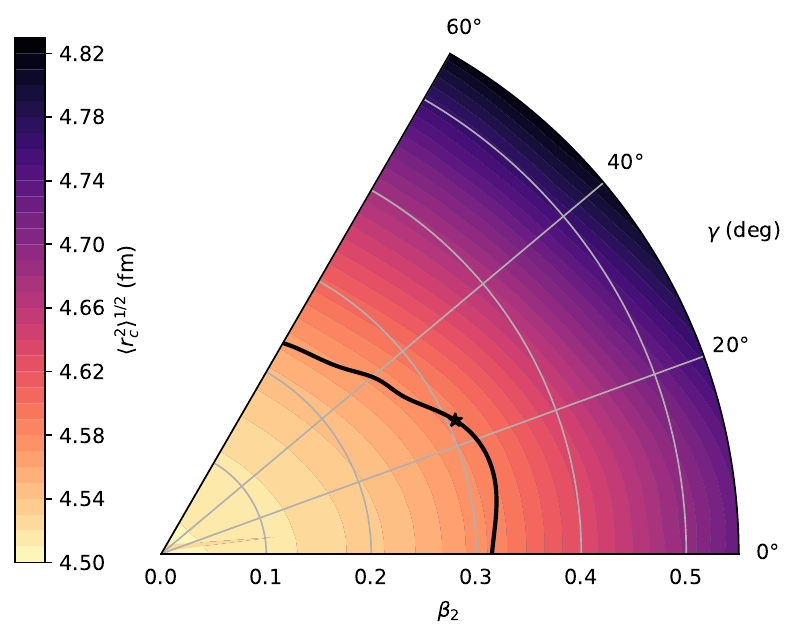}
    \caption{  Root-mean-square charge radius $R_c = \langle r_c^2 \rangle^{1/2}$ of $^{112}$Ru across the $\beta-\gamma$ plane, as calculated for BSkG4. The black line and black star are as in Fig.~\ref{fig:pes}.
    }
    \label{fig:radiuspes}
\end{figure}

We can illustrate the consequences of these effects for $^{112}$Ru using BSkG4~\footnote{We show in the supplementary material that this is a representative choice.}: Fig.~\ref{fig:pes} and Fig.~\ref{fig:radiuspes} show the system's total energy and charge radius, respectively, as a function of $\beta_2$ and $\gamma$. The black lines indicate the total deformation that minimizes the energy as a function of the triaxiality angle, illustrating that $\beta_2$ varies in a non-trivial way along this path. In particular, the values of $\beta_2$ at the lowest prolate and oblate axial configurations and at the triaxial minimum are quite different. Even more important is that $R_c$ has a visible $\gamma$ dependence. At fixed $\beta_2$, the calculated $R_c$ increases with $\gamma$ by more than 0.03\,fm. This is the opposite of what is expected for a purely quadrupole-deformed liquid drop: at $\beta_2 \simeq 0.3$, this model predicts a decrease by about 0.01\,fm when going from prolate to oblate shape. Part of the difference between the two models can be attributed to an increase of higher-order deformations with $\gamma$ in the EDF model and the other part results from effects not captured by the liquid drop model.  

The details of shell effects vary from one model to the next, which naturally means that the effect of triaxial deformation on the nuclear charge radius becomes model-dependent, as demonstrated in the supplemental material. Even if this were not the case, a single value for the charge radius cannot determine a unique point on the $(\beta_2, \gamma)$ plane. 
Measuring R$_c$ will not allow us to determine whether or not a given nucleus exhibits triaxial deformation. This does not mean the effect is unimportant: Fig.~\ref{fig:bskg} shows that calculations that account for this degree of freedom perform markedly better than those restricted to axial symmetry -- the latter also disagree with the available spectroscopic information for these isotopes. 
When taken together, all of these considerations paint a coherent picture of ground state triaxial deformation; it is only by accounting for this deformation that the BSkG models arrive at their excellent and simultaneous description of masses~\cite{hukkanen2023,hukkanen2023a}, spectroscopy~\cite{stryjczyk2024} and charge radii in this region of the nuclear chart.

\paragraph{Conclusions}
The installation of the new ATLANTIS beamline allowed us to perform isotope shift measurements of neutron-rich ruthenium produced from $^{252}$Cf fission at CARIBU. The highly efficient RFQ cooler and buncher and a magnesium-based charge-exchange cell compensated for the weak fission rate. We extracted the differential rms charge radii of the nine radioactive isotopes $^{106-114}$Ru from the isotope shifts.

There is significant evidence that these isotopes of Ru are triaxial, and only the inclusion of triaxial deformation into the calculations allows us to describe the experimental data while being consistent with all known information about the nuclei in question. The deformation impacts calculated charge radii through shell effects, contrary to what is expected from liquid drop modeling.
 
Earlier studies with the BSkG series already pointed to the relevance of triaxial deformation for the behavior of binding energies and isomeric states in this region~\cite{hukkanen2023,hukkanen2023a,stryjczyk2024}. Here, we show that this combined description extends to charge radii: the BSkG$x$ describe the new data remarkably well. A consistent picture is emerging: triaxial deformation leaves a fingerprint on essentially all properties of nuclei in this region of the nuclear chart; our work illustrates how spontaneous symmetry breaking in strongly correlated systems not only impacts spectroscopic properties but also bulk observables.

\paragraph*{Acknowledgments}
A share of the research work described herein originates from R\&D carried out in the frame of the FAIR Phase-0 program of LASPEC/NUSTAR. W. R. is a Research Associate of the F.R.S.-FNRS. This work was supported by the Deutsche Forschungsgemeinschaft (DFG, German Research Foundation) – Project-ID 279384907 – SFB 1245 and NSF Grant No. PHY-21-11185, by the U.S. Department of Energy, Office of Nuclear Physics, under Contract No. DE-AC02-06CH11357, with resources of ANL’s ATLAS facility, an Office of Science User Facility, and from the German Federal Ministry for Education and Research under Grants 05P15RDFN9 and 05P21RDFN1.This work was supported by the Fonds de la Recherche Scientifique (F.R.S.-FNRS) and the Fonds Wetenschappelijk Onderzoek - Vlaanderen (FWO) under the EOS Projects nrs. O022818F and O000422F. This research benefited from computational resources made available on the Tier-1 supercomputer Lucia of the Fédération Wallonie-Bruxelles, infrastructure funded by the Walloon Region under the grant agreement nr 1117545. Further computational resources have been provided by the clusters of the Consortium des Équipements de Calcul Intensif (CÉCI), funded by the F.R.S.-FNRS under Grant No. 2.5020.11 and by the Walloon Region. 
\clearpage
\section{Supplementary material}
\subsection{Details of the fit procedure}
All spectra were fit with a $\chi^2$ minimization routine; The absolute count rates (including background) for all spectra are high enough to follow a Poisson distribution. The peak shape is a symmetric Voigt profile. The peak width is fixed to the average $\gamma=32.8\,(1.3)$\,MHz of a series of stable beam measurements, and $\sigma$ is left free for each spectrum, yielding values around $15$\,MHz. The isotope shift proved to be insensitive to the exact choice of $\gamma$, even when a pure Lorentz or Gaussian profile was enforced.
\subsection{Details of the King plot procedure}
\begin{figure}
    \centering
    \includegraphics[width=1\linewidth]{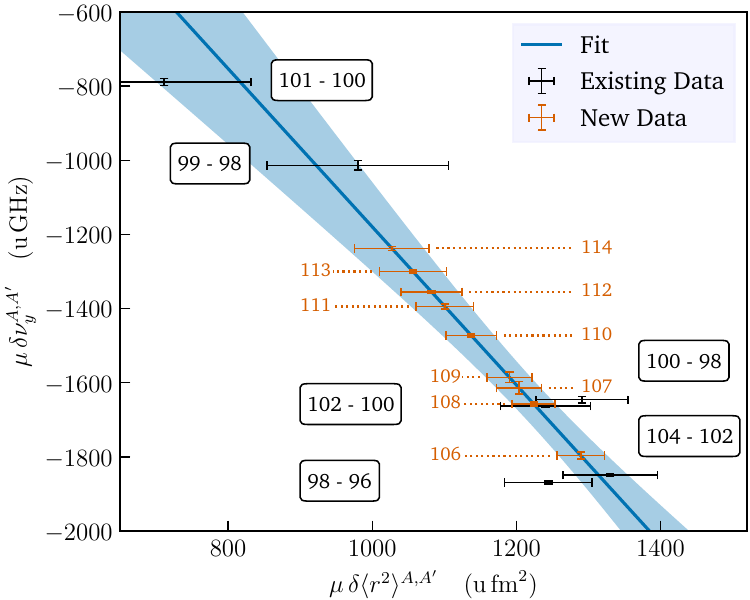}
    \caption{A King fit of the (modified) nuclear parameter $\Lambda^{A,A'}_\mu$ and (modified) isotope shifts in stable nuclei, shown in black. The new $\Lambda^{102,106-114}$, shown in yellow, can be extracted by using the isotope shifts presented in this letter.}
              \label{fig:king}
\end{figure}
The King plot, shown in Fig.~\ref{fig:king}, was performed using isotope shifts $\delta\nu^{A,A'}$ of stable isotopes, which were measured with high precision at ATLANTIS. Their values agree well with previous experiments \cite{Forest2014} but have higher precision.
We used the corresponding nuclear radius parameters $\Lambda^{A,A'}_\mu$ given in \cite{Schopper2004}; They are calculated from Barrett equivalent radii $R^\mu_{k\alpha}$ found with muonic atom spectroscopy. For the calculation, the ratios 
\begin{equation}
    V_n= R^\mu_{k\alpha} \langle r^n \rangle^{-\frac{1}{n}}~\textrm{($n=2,4,6$)}
\end{equation}
of the radial moments for $^{100}$Ru were used for all isotopes, and their values 
\begin{equation}
\begin{aligned}
    V_2 &= 1.2802 \\
    V_4 &= 1.1911 \\
    V_6 &= 1.1234 \\ 
\end{aligned}
\end{equation}
are given without uncertainty in~\cite{Schopper2004}. The nuclear radius parameters
\begin{equation}
    \Lambda^{A,A'}_\mu = \sum_{i=1,2,3;\,\,n=2i}C_i\left[\left(\frac{R^\mu_{k\alpha}}{V_n}\right)_{A'}^2 - \left(\frac{R^\mu_{k\alpha}}{V_n}\right)_{A}^2\right]
\end{equation}
can be calculated with the help of Seltzer coefficients \cite{Seltzer1969} 
\begin{equation}
\begin{aligned}
C_1 &=  1\,\textrm{fm}^{-2} \\
C_2 &= -5.39\cdot10^{-4}\,\textrm{fm}^{-4} \\
C_3 &=  1.78\cdot10^{-6}\,\textrm{fm}^{-6}\quad\textrm{.}
\end{aligned}
\end{equation}
In contrast to the King fit performed in~\cite{Schopper2004}, we introduced an additional 0.02\,\% uncertainty on each $V_n^A$ to incorporate the possibility of an isotope dependence of the radial moments, in the same magnitude as observed in the neighboring elements molybdenum and palladium. The resulting values and their uncertainty used here can be found in Tab.~\ref{tab:king_input}. With the additional uncertainty in $V_n^A$, the data agrees well with the linear King fit~($\chi_{\nu}^2=1.0$), which yields 
\begin{equation}
\begin{aligned}
F&=-2.13\,(46)\,\textrm{\,GHz\,fm}^{-2}   \\
K_\infty&=954\,(564)\,\textrm{\,u\,GHz} 
\end{aligned}
\end{equation} with the covariance matrix 
\begin{equation}
\begin{aligned}
&\mathbf{\text{Cov}} =
\begin{bmatrix}
\textrm{Var}\left(F\right) & \textrm{Cov}\left(F,K\right) \\
\textrm{Cov}\left(K,F\right) & \textrm{Var}\left(K\right)
\end{bmatrix}\\  =
&\begin{bmatrix}
318\,088\,\textrm{GHz}^{2}\,\textrm{fm}^{-4} & -256.567\,\textrm{GHz}^{2}\,\textrm{u}\,\textrm{fm}^{-2} \\
-256.567\,\textrm{GHz}^{2}\,\textrm{u}\,\textrm{fm}^{-2} & 0.20958\,\textrm{GHz}^{2}\,\textrm{u}^2
\end{bmatrix}
\end{aligned}
\end{equation}
which allows the uncertainties of the King plot to be propagated to the nuclear radius parameters of the radioactive isotopes. This is visualized with the red data points in Fig.~\ref{fig:king}. We also investigated explicitly adding an isotope dependence of $V_n$, similar to that in molybdenum and palladium, to the calculation of $\Lambda^{A,A'}_\mu$. The resulting new charge radii systematically shift within the uncertainty given in this Letter. The uncertainty in the Field shift factor is comparatively large; This could be resolved by more accurate radii measurements of the stable isotopes, or, potentially, with advanced atomic theory calculations, which have demonstrated the required accuracy, e.g., in indium.~\cite{Sahoo2020}
\begin{table}
\centering
\ra{1.2}
\caption{The input values for the King fit. The isotope shift $\delta\nu^{A,A'}$ was measured with high precision at ATLANTIS. The nuclear radius parameter is extracted from \cite{Schopper2004} and uses Barrett equivalent radii from muonic atom spectroscopy, ratios of the radial moments, and Seltzer's coefficients given in the same publication.}

\begin{tabularx}{0.49\textwidth}{rXrXD..{3.1}lXD..{1.4}l}
\tls
\toprule[\lightrulewidth]\toprule[\lightrulewidth]

$A$       && $A'$    && \multicolumn{2}{c}{$\delta\nu^{A,A'}$}     && \multicolumn{2}{c}{$\Lambda^{A,A'}_\mu$}    \\
        &&       && \multicolumn{2}{c}{(MHz)}                   && \multicolumn{2}{c}{(fm)}         \\
\colrule
96      && 98    && -397.9       & (0.8)                         && 0.2649      & (129)          \\
98      && 100   && -336.5       & (1.8)                         && 0.2639      & (131)          \\
100     && 102   && -326.8       & (0.8)                         && 0.2438      & (123)          \\
102     && 104   && -349.6       & (0.4)                         && 0.2515      & (124)          \\
98      && 99    && -104.8       & (1.3)                         && 0.1013      & (130)          \\
100     && 101   &&  -78.4       & (0.9)                         && 0.0707      & (120)          \\
\bottomrule[\lightrulewidth]\bottomrule[\lightrulewidth]
\end{tabularx}
\label{tab:king_input}
\end{table}

\subsection{Characterizing the nuclear shape}

The nuclear shape can either be characterized through a parameterization of its surface or its (volume) multipole moments. The former of these two concepts is used for macroscopic-microscopic models. One possibility, but by far not the only one \cite{hasse1988}, is to expand the surface $R(\vartheta, \varphi)$ into spherical harmonics $Y_{\ell m} (\vartheta, \varphi)$ with coefficients 
$\beta_{\ell m}^{\rm LD}$
\begin{align}
\label{eq:betalm:LDM}
R(\vartheta, \varphi)
& = R_d \big( \{ \beta_{\ell m}^{\rm LD} \} \big)
    \Bigg[ 1 + \sum_{\ell m} \beta_{\ell m}^{\rm LD} \, Y_{\ell m} (\vartheta, \varphi) \Bigg]
    \, .
\end{align}
In such models, the surface of the proton and neutron density distribution is assumed to be the same. To ensure that the nuclear volume remains constant when changing the deformation, the reference radius $R_d$ is not exactly equal to the radius of the sharp surface of a spherical liquid drop $R_0^{\rm LD} = r_0 A^{1/3}$, but has itself a deformation dependence that can be expressed through a multivariate 
polynomial in the $\beta_{\ell m}^{\rm LD}$ \cite{hasse1988}. Whatever the actual parameterization used for the surface, the shape parameters defining $R(\vartheta, \varphi)$ are the variational parameters of this kind of approach.

The other possibility to characterize nuclear shapes is to calculate the expectation values of the spherical multipole moments 
$\hat{Q}_{\ell m} \equiv \hat{r}^{\ell} \, \hat{Y}_{\ell m} (\vartheta, \varphi)$
of a given density distribution $\rho(\bvec{r})$
\begin{align}
\langle \hat{Q}_{\ell m} \rangle 
& = \int \, \mathrm{d}^3 r \, r^\ell \, Y_{\ell m} (\vartheta, \varphi) \, \rho(\bvec{r}) \, ,
\label{eq:defQLM}
\end{align}
where $\vartheta$ and $\varphi$ are the polar and azimuthal angles of the radius vector $\bvec{r}$ and $r$ its radial component. 

Such multipole moments are the natural choice to characterize the shapes of nuclei calculated with energy density functional (EDF) methods. With the symmetries chosen for the EDF calculations reported here, all multipole moments $\langle \hat{Q}_{\ell m} \rangle$ are real and can take finite values only when $\ell$ and $m$ are both even with $\langle \hat{Q}_{\ell m} \rangle = \langle \hat{Q}_{\ell -m} \rangle$. Note that in EDF models, the individual multipole moments of the proton and neutron densities are generally slightly different.

The multipole moments are most transparently discussed in terms of dimensionless deformation parameters $\beta_{\ell m}$ defined as
\begin{align}
\beta_{\ell m} 
& = \frac{4 \pi } {  3 R_0^{\ell} A }  \langle \hat{Q}_{\ell m} \rangle \, ,
\label{eq:betalm}
\end{align}
where we use the convention that $R_0 = 1.2 A^{1/3} \, \mathrm{fm}$ for the normalization as done in tabulations of transition quadrupole moments \cite{Raman.2001}. Sometimes, the calculated 
root-mean-square radius of the same density distribution is used instead in the literature to discuss deformations predicted by models. For atomic nuclei, the dominant multipole deformations are the two components of the quadrupole tensor $\beta_{20}$ and $\beta_{22}$ in the nucleus' major axis system in which $\beta_{21} = \beta_{2-1} = 0$ and $\beta_{22} = \beta_{2-2}$ is real. These are often translated to the more intuitive \emph{total quadrupole deformation} $\beta_2$ and the \emph{triaxiality angle} $\gamma$
\begin{align}
\label{eq:beta2}
\beta_2
& = \sqrt{\beta_{20}^2 + 2\beta_{22}^2} \, ,
    \\
\label{eq:gamma}
\gamma
& = \mathrm{atan} \bigg( \sqrt{2} \, \frac{\beta_{22}}{\beta_{20}} \bigg)  \, .
\end{align}
The latter differentiates between non-axial shapes, $\gamma \in (0^{\circ}, 60^{\circ})$,  and those with one rotational symmetry axis: prolate ($\gamma = 0^{\circ}$) or oblate ($\gamma = 60^{\circ}$) shapes.

When calculating the multipole moments~\eqref{eq:betalm} of a liquid-drop with a sharp surface defined through Eq.~\eqref{eq:betalm:LDM}, one finds that at small deformation, the values of $\beta_{\ell m}$ and $\beta_{\ell m}^{\rm LD}$ are of similar size, but they are not equivalent. In particular, any given $\beta_{\ell m}$ is, in general, a multivariate polynomial in several shape degrees of freedom 
$\beta_{\ell' m'}^{\rm LD}$ \cite{hasse1988}.

\subsection{The BSkG$x$ models}

The charge radii reported in the main text are calculated with the microscopic BSkG models~\cite{Scamps.2021,Ryssens.2022,Grams.2023,Grams.2025}. These are \textit{global} models in the sense that they aim to simultaneously describe the largest possible range of observables for low-lying states of all nuclei with the same EDF and within the same approach. One of the ultimate goals of this line of research is to have a model that provides all relevant data for $r$-process nucleosynthesis simulations~\cite{Scamps.2021}. The BSkG$x$ models are based on self-consistent HFB calculations with Skyrme-type EDFs to which some corrections are added, most importantly corrections for rotational and vibrational motion. 

For the majority of Skyrme parametrizations, the coupling constants are adjusted to properties of nuclear matter and a few selected spherical nuclei. This is different for the BSkG$x$, which are adjusted to all nuclei across the nuclear chart, including deformed ones. The MOCCa code~\cite{Ryssens2016} that represents the single-particle wave functions on a three-dimensional mesh in coordinate space has been used for this, allowing all nuclei to take an energetically optimal shape that might be spherical, axial, or triaxial during the parameter adjustment. With this, the BSkG$x$ models are particularly well suited to study the evolution of shapes in the region of the nuclear chart addressed in the main text. However, no information on nuclear shape entered the objective function of the parameter adjustment, such that all multipole moments are \emph{predictions} of the model. Furthermore, the energetically optimal configurations we construct might have non-vanishing $\beta_{\ell m}$ for any combination of $\ell$ and $m$ that is not restricted by symmetries; in particular our calculations that include triaxial deformation are not only sensitive to $\beta_{22}$, but also
to $\beta_{42}, \beta_{44}, \beta_{62}, \ldots$.

Even if motivated by astrophysical applications, the BSkG models are at the forefront of the global modeling of nuclear structure: aside from other attractive properties not relevant to this study, BSkG1-4 feature root-mean-square (rms) deviations of less than 750 keV for all known nuclear masses and less than 0.028 fm for hundreds of known rms charge radii~\cite{Scamps.2021,Ryssens.2022,Grams.2023,Grams.2025}. 

The BSkG models describe each nucleus with a single Bogoliubov quasiparticle vacuum that has a well-defined shape. The ground state is determined variationally for each nucleus as the Bogoliubov 
quasiparticle state that yields the largest binding energy, thereby fixing the nuclear shape and radius. All calculations reported here were performed with the MOCCa code~\cite{Ryssens2016} under the exact numerical conditions of the BSkG$x$ parameter adjustment.

We mention that all results for odd-mass isotopes were obtained through fully self-consistent calculations that minimize the total energy of a blocked quasiparticle minimum. For each odd nucleus, such calculations were run for a set of suitably selected candidates for blocked quasiparticles in order to identify  the one-quasiparticle configuration with the lowest total energy \emph{after} convergence is reached. Beginning with BSkG2, these calculations include the contributions of so-called time-odd terms~\cite{Ryssens.2022,Grams.2023}, whereas for the earlier BSkG1 the simpler equal filling approximation that this model was adjusted with~\cite{Scamps.2021}, is used.

\subsection{Calculating charge radii}

Nucleons are composite objects; therefore, the nuclear charge density has to be calculated, including corrections for the internal charge distribution of the nucleons. Broadly speaking, two strategies to calculate charge radii coexist in the literature on EDF models. One consists of adding corrections to the mean-square radius of the point-proton density, and the other is to calculate all electromagnetic observables from a suitably defined local charge density. The BSkG$x$ use the latter, which has the advantage that all electromagnetic observables can be consistently calculated, which for the BSkG$x$ also includes the Coulomb potential and the electric energy. Various possibilities exist to define the charge density used in the literature. The BSkG$x$ include the contributions from the internal charge density $f_q(\bvec{r})$ of protons and neutrons
\begin{align}
\rho_c(\bvec{r})
& = \int \! \mathrm{d}^3 r' 
    \big[ f_p(\bvec{r} -\bvec{r}') \, \rho_p(\bvec{r}')
         +f_n(\bvec{r} -\bvec{r}') \, \rho_n(\bvec{r}') 
    \big] \, .
\end{align}
The charge form factor of protons is modeled with a single normalized Gaussian $f_p(\bvec{r}) = e^{-\bvec{r}^2/\mu_p} / (\pi \mu_p)^{3/2}$ whose range parameter $\mu_p$ is determined from the rms charge radius of the proton through $r_p = \sqrt{3\mu_p/2} = 0.895 \, \text{fm}$ \cite{Scamps.2021}. By contrast, the charge form factor of the neutrons is modeled as the sum of two normalized Gaussians of opposite sign
$f_n(\bvec{r}) = e^{-\bvec{r}^2/\mu_+} / (\pi \mu_+)^{3/2}
-e^{-\bvec{r}^2/\mu_-} / (\pi \mu_-)^{3/2}$ with range parameters
$\mu_{+} = r_{+}^2 = 0.387 \, \text{fm}^2$ and
$\mu_{-} = r_{-}^2 = 0.467 \, \text{fm}^2$\cite{Scamps.2021}, which corresponds to a neutron with net charge zero, but a ms charge radius of $\tfrac{3}{2} \big( \mu_{+} - \mu_{-} \big) = -0.12 \, \text{fm}^2$. Additional corrections for relativistic effects are also sometimes added to the charge density but have been neglected for the BSkG$x$.

From this charge density, the mean-square charge radius of the nucleus 
is then simply calculated as 
\begin{align}
\langle r^2_c \rangle 
& = \frac{1}{Z} \int \! \mathrm{d}^3r \, r^2 \, \rho_c(\bvec{r}) \, .
\end{align}
The advances made in the evolution from BSkG1 to BSkG4 are not directly 
related to the description of nuclear charge radii. Thus, we have no prior reason to expect markedly better predictions from any of them for this study. We employ all four to have an indication of the systematic uncertainties of the model.

\subsection{Impact of triaxial shapes on charge radii}

Assuming that the nuclear surface \eqref{eq:betalm:LDM} only has a quadrupole deformation and transforming $\beta_{2 0}^{\rm LD}$ and $\beta_{2 2}^{\rm LD}$ of Eq.~\eqref{eq:betalm:LDM} to $\beta^{\rm LD}_{2}$ and $\gamma^{\rm LD}$ along the lines of Eqs.~\eqref{eq:beta2} and \eqref{eq:gamma} \cite{hasse1988}, the ms charge radius of a triaxial liquid drop is given by
\begin{align}
\label{eq:LDMexpression}
\langle r^2_c \rangle  
& = \frac{3}{5} \big(R_0^{\rm LD} \big)^2 
    \bigg[ 1 
           + \frac{5}{4\pi} (\beta^{\rm LD}_{2})^2 
           \nonumber \\
& \qquad \qquad
           + \frac{5}{4\pi} \sqrt{\frac{5}{4\pi}} \, \frac{10}{21} \, 
            \left(\beta_2^{\rm LD}\right)^3 \cos(3\gamma^{\rm LD}) 
    \bigg] \, ,
\end{align}
where $\tfrac{3}{5} \big( R_0^{\rm LD} \big)^2$ is the ms charge radius of a spherical liquid drop of charge $Z$ and sharp radius $R_0^{\rm LD} = r_0 A^{1/3}$. The expression it multiplies contains the correction for volume conservation, i.e.\ the difference between $R_0^{\rm LD}$ and $R_d$ of Eq.~\eqref{eq:betalm:LDM}, that actually cancels half of the leading-order contribution $\sim (\beta^{\rm LD}_{2})^2$ from deformation to the ms radius and therefore cannot be neglected.

\begin{figure*}
\includegraphics[width=.3\textwidth]{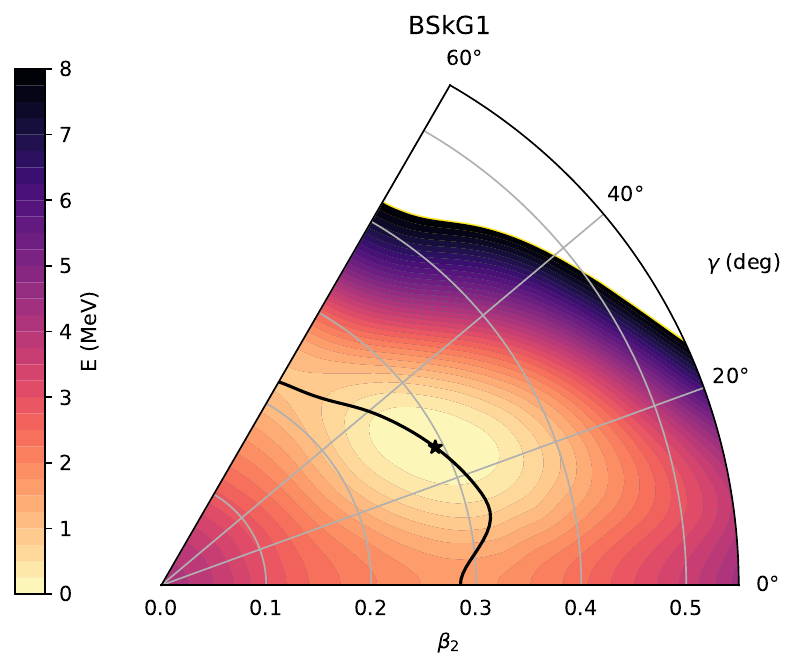}
\includegraphics[width=.27\textwidth]{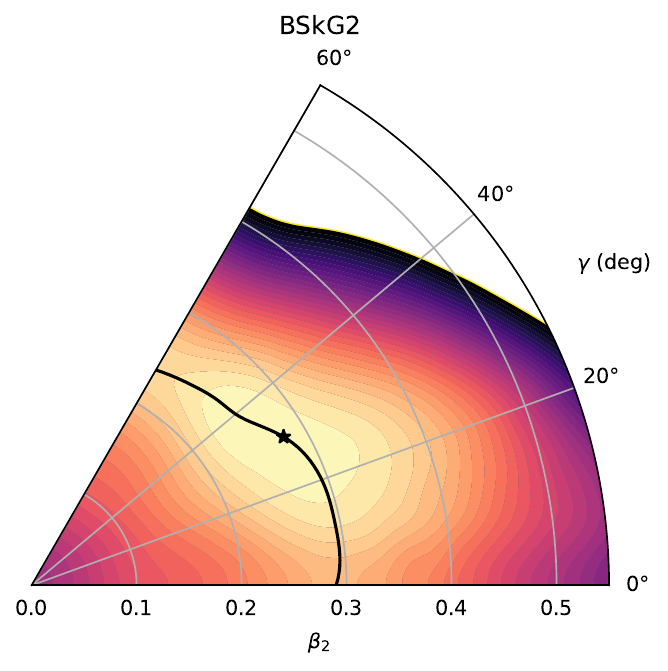}
\includegraphics[width=.27\textwidth]{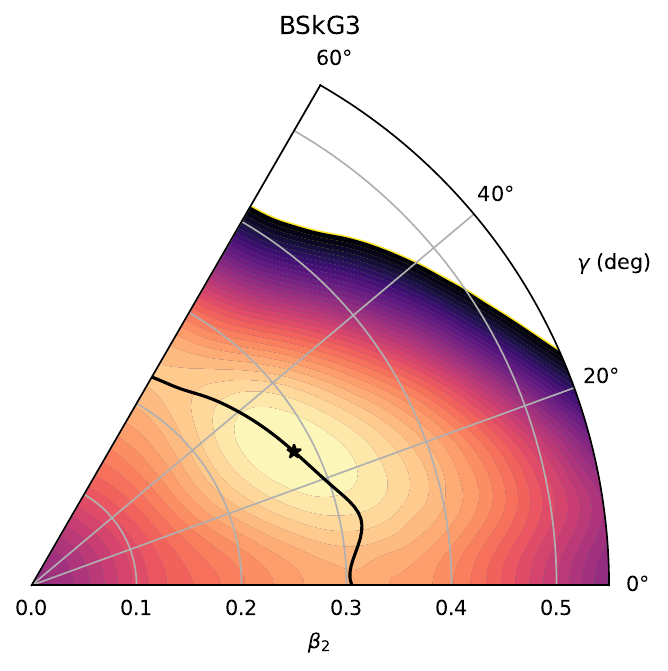}
\caption{Normalized total energy of $^{112}$Ru as a function of $\beta_{2}$ and $\gamma$ as obtained with BSkG1 (left), BSkG2 (middle) and BSkG3 (right). The black lines indicate in each case the minimal energy as a function of $\gamma$; a black star indicates the global energy minimum. }
\label{fig:supp:pes}
\end{figure*}

\begin{figure*}
\includegraphics[width=.3\textwidth]{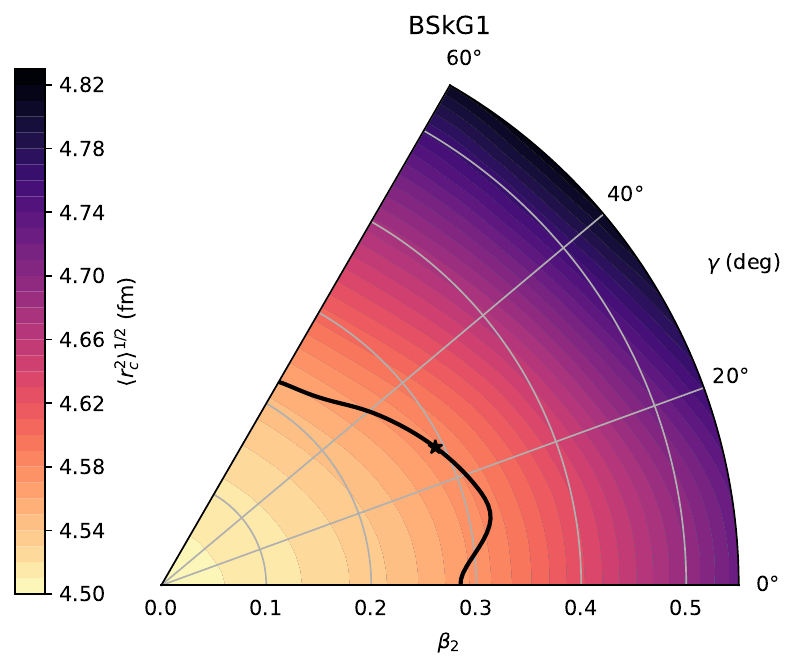}
\includegraphics[width=.27\textwidth]{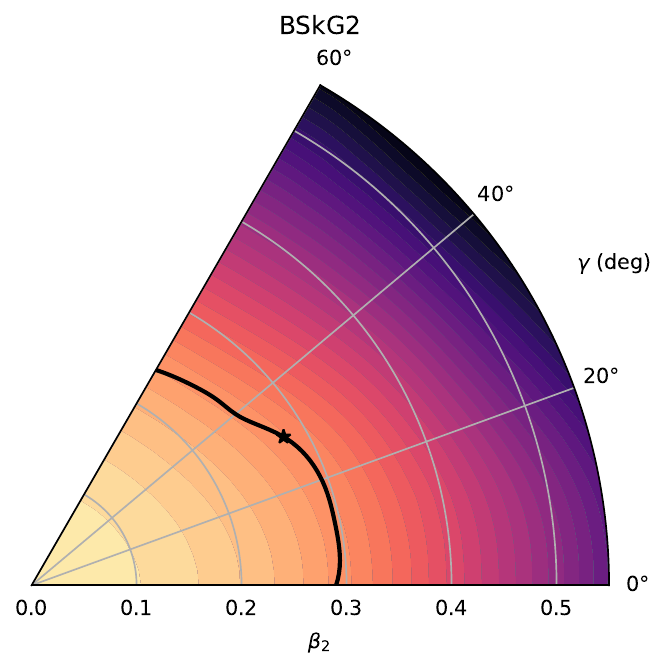}
\includegraphics[width=.27\textwidth]{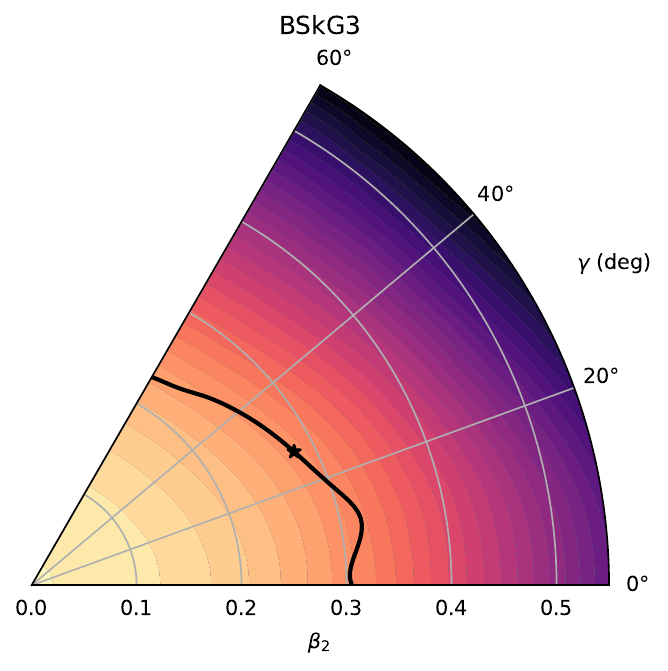}
\caption{ Root-mean-square charge radius of $^{112}$Ru across the $\beta - \gamma$ plane, as calculated for BSkG1 (left), BSkG2 (middle ) and BSkG3 (right). The black lines and black stars are as in Fig.~\ref{fig:supp:pes}.}
\label{fig:supp:radiuspes}
\end{figure*}

Note that the expression in the original paper on the possible role of 
triaxiality on nuclear radii \cite{grechukhin1960} is incorrect.
Using that $\cos(\gamma) \big[ 1 - 4 \sin^2 (\gamma) \big] = \cos(3\gamma)$, expression~\eqref{eq:LDMexpression} is equivalent to the first equation in the Erratum \cite{grechukhin} to Ref.~\cite{grechukhin1960}. In the Erratum \cite{grechukhin}, the author also gives a simpler approximate expression that, unfortunately, contains a typographical error. A correct rewrite of the approximate expression has later been given in Ref.~\cite{hilberath1992}. This approximation amounts to replacing the factor $\sqrt{\frac{5}{4\pi}} \, \frac{10}{21} \simeq 0.30037$ in the third term on the r.h.s.\ of Eq.~\eqref{eq:LDMexpression} by $\frac{3}{10}$.

\rule{0mm}{4mm} Assuming $\beta_{2}^{\rm LD}$ to be fixed and small, varying $\gamma^{\rm LD}$ will affect the radius of a liquid drop only at subleading order. Going from $\gamma^{\rm LD} = 0$ to 60 degrees then actually smoothly interpolates between the positive subleading correction found for axial prolate shapes and the negative subleading correction found for axial oblate shapes. With this, no distinct quantitative or qualitative signature of triaxiality affects the evolution of charge radii $\langle r^2_c \rangle$ in an obvious manner.  

To complicate matters further, the correction from triaxiality is competing with the contributions from higher-order shape deformations not included in Eq.~\eqref{eq:LDMexpression}. These all bring a leading-order correction  $5/(4\pi) \, \big(\beta_{\ell}^{\rm LD}\big)^2$ with
\begin{equation}
\big( \beta_{\ell}^{\rm LD}\big)^2
= \sum_{m = -\ell}^{\ell} \big| \beta_{\ell m}^{\rm LD} \big|^2 
\end{equation}
to the ms charge radius \cite{hasse1988} plus numerous additional multivariate subleading corrections. Even if the higher-order deformation parameters remain, in general, smaller than $\beta_{2}^{\rm LD}$, their contribution to $\langle r^2_c \rangle$ can often be as large or larger than the impact of triaxiality on the subleading correction in Eq.~\eqref{eq:LDMexpression}.

And indeed, inspection of Fig.~4 in the main text reveals a direct $\gamma$-dependence of the rms charge radius that is the opposite of the one expected from Eq.~\eqref{eq:LDMexpression}: the BSkG 
parameterizations predict that at equal $\beta_2$ the radius grows with increasing $\gamma$. This illustrates the limitations of interpreting the $\gamma$-dependence of $\langle r^2_c \rangle$ based on the schematic liquid-drop estimate \eqref{eq:LDMexpression} that is restricted to quadrupole deformation.

The EDF calculations indicate, for example, that the hexadecapole moments $\beta_{40}, \beta_{42}$ and $\beta_{44}$ of $^{112}$Ru vary across the $\beta_2$-$\gamma$ plane. Even more importantly, the lowest energy found at different gamma is located at different total quadrupole deformation $\beta_2$, which through the leading-order correction in Eq.~\eqref{eq:LDMexpression} affects charge radii more than the impact of changing $\gamma$ in the sub-leading term.

The impact of triaxiality on the nuclear charge radius is not unique to BSkG4. We show in Fig.~\ref{fig:supp:pes} the potential energy surfaces obtained with BSkG1-2-3. These are all qualitatively similar to the potential energy surface obtained with BSkG4 (Fig.~3 in the main text), with a rather broad triaxial minimum near $\beta_2 \sim 0.285$ and $\gamma \sim 27^{\circ}$. In all cases, this global minimum is more than 750 keV lower in energy than the oblate and prolate saddle points. Although the location of the triaxial minimum is quite consistent across models, the quadrupole deformation of the axial saddle points - especially the prolate one - is somewhat model-dependent. We point out in particular that the triaxial minimum for BSkG1 has larger $\beta_{2}$ than both the prolate and oblate saddle points. This example proves that the total quadrupole deformation of a triaxial minimum is not necessarily in between that of the axially symmetric saddle points, even if it is the case for the PES of this nucleus for BSkG2, BSkG3 and BSkG4. Fig.~\ref{fig:supp:radiuspes} shows the corresponding rms charge radii in the $\beta - \gamma$ plane. All surfaces are qualitatively 
similar to Fig.~4 in the main text, indicating a dependence of $R_c$ on $\gamma$ that does not match what is expected from a purely quadrupole-deformed liquid drop.

\begin{figure}
    \centering
    \includegraphics[width=0.45\textwidth]{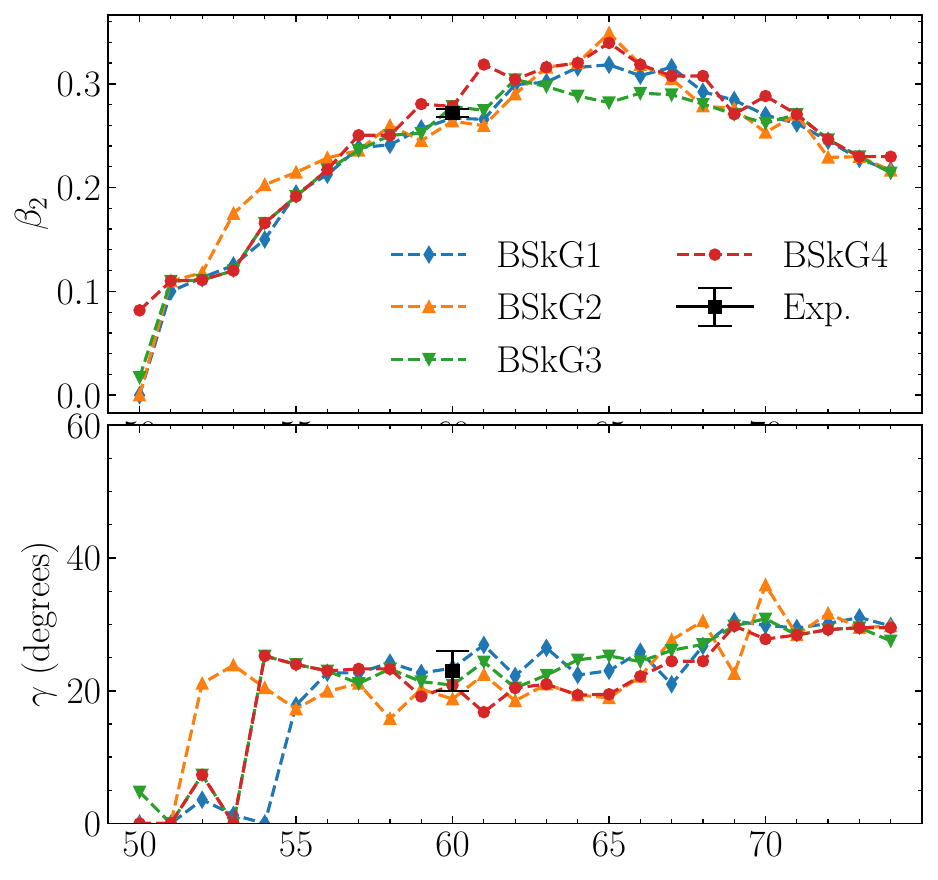}
    \caption{Evolution of the quadrupole deformation of Ru isotopes as obtained with the BSkG$x$. 
    Top panel: total quadrupole deformation $\beta_2$. Bottom panel: triaxiality angle $\gamma$. The experimental values for $\beta_2$ and $\gamma$ are taken from \cite{nudat} and \cite{Srebrny.2006}, respectively.}
    \label{fig:supp:evolution}
\end{figure}

Finally, we establish that (i) all BSkG$x$ essentially agree on the evolution of quadrupole deformation along the Ru chain and that (ii) our example nucleus $^{112}$Ru is not an outlier. Fig.~\ref{fig:supp:evolution} shows the evolution of $\beta_2$ (top panel) and $\gamma$ (bottom panel) for the Ru isotopic chain; in the range $N=56$ to $72$ all BSkG models predict a nearly constant 
value of $\gamma$ and a smoothly evolving $\beta_2$ roughly between 0.2 ($N = 56$) and 0.3 ($N=66$). The sole experimental data point concerns the COULEX data on $^{104}$Ru of Ref.~\cite{Srebrny.2006} that was already mentioned in the main text. Only for the lightest isotopes we consider here, there is a significant model difference: BSkG1 predicts no triaxiality below $N < 55$, while the corresponding bounds for BSkG2, BSkG3 and BSkG4 are $N < 52$, $N < 54$ and $N < 54$ respectively.

\bibliographystyle{prl}
\bibliography{ru_spec}

\end{document}